\documentclass[prd,aps,showpacs,11pt,
nofootinbib,%
tightenlines,
notitlepage,
]{revtex4-1}

\usepackage{graphicx}
\usepackage{amsmath}
\usepackage{amssymb}
\usepackage{hyperref} 
\usepackage{colortbl}
\usepackage{soul}
\sethlcolor{yellow}

\newcommand{\be}{\begin{equation}}
\newcommand{\ee}{\end{equation}}
\newcommand{\bea}{\begin{eqnarray}}
\newcommand{\eea}{\end{eqnarray}}

\newcommand{\6}{\partial }

\newcommand{\MKK}{M_{\rm KK}}

\newcommand{\Tr}{{\rm Tr}\,}

\newcommand{\AD}{$\overline{\text{D8}}$}

\def\PDG{\cite{PDG18}}

\renewcommand{\hl}[1]{#1}
\renewcommand{\st}[1]{}

\begin{document}

\title{Pseudoscalar transition form factors and the hadronic light-by-light contribution to the 
anomalous magnetic moment of the muon\\
from holographic QCD
}


\author{Josef Leutgeb}
\author{Jonas Mager}
\author{Anton Rebhan}
\affiliation{Institut f\"ur Theoretische Physik, Technische Universit\"at Wien,
        Wiedner Hauptstrasse 8-10, A-1040 Vienna, Austria}

\date{\today}

\begin{abstract}
We revisit the predictions for the pseudoscalar-photon transition form factors in 
bottom-up and top-down holographic QCD models which only use the pion decay constant and the $\rho$ meson mass as input.
We find remarkable agreement with the available experimental data for the single-virtual $\pi^0$ form factor
that have recently been extended to lower momenta by BESIII, down to 0.3 GeV$^2$. The bottom-up models moreover turn out to be roughly 
consistent with recent experimental results obtained by BaBar for the double-virtual $\eta'$ form factor at large momenta
as well as with dispersion theory results and
a recent lattice extrapolation for the double-virtual $\pi^0$ form factor. 
Calculating the 
pion pole contribution to the hadronic light-by-light scattering in the anomalous magnetic moment of the muon, 
we find that the bottom-up models in question span the range
\hl{$a_\mu^{\pi^0}=6.1(4)\cdot 10^{-10}$}, which is \st{about 10\%}\hl{somewhat} lower than estimated previously
by approximating these holographic predictions through simple interpolators,
\hl{and in remarkably good agreement with recent results based on a dispersive approach
or lattice simulations.}
\end{abstract}

\maketitle

\section{Introduction}

The current world average for the experimental value of the anomalous magnetic moment of the muon
\cite{Jegerlehner:2009ry,Jegerlehner:2017gek}, $a_\mu=(g_\mu-2)/2$, 
which is dominated by the final result of the E821 collaboration at Brookhaven National Laboratory obtained 15 years ago,
reads \PDG
\be
a_\mu^{\rm exp.}=(11659209.1 \pm 6.3) \times 10^{-10},
\ee
with new experiments at FERMILAB and J-PARC aiming at an even more precise determination.
According to the recent update in Ref.~\cite{Davier:2019can}, the Standard Model prediction is 3.3 standard deviations below this value, at%
\footnote{See Ref.~\cite{Keshavarzi:2018mgv} and \cite{Jegerlehner:2017lbd} for other recent updates which posited even higher discrepancies
at 3.7 and 4.1 standard deviations, respectively.}
\be\label{amuSM}
a_\mu^{\rm theory}=(11659183.0 \pm 4.8) \times 10^{-10},
\ee
with very precisely determined contributions from QED \cite{Aoyama:2012wk} and electroweak effects \cite{Czarnecki:2002nt,Gnendiger:2013pva}.
The uncertainty is almost entirely due to strong-interaction contributions \cite{Davier:2019can,Kurz:2014wya,Colangelo:2014qya}, with the
hadronic light-by-light (HLBL) scattering contribution having an estimated error of about $\pm 3\times 10^{-10}$
\cite{Prades:2009tw,Jegerlehner:2017gek}. The latter is dominated by the exchange of a pseudoscalar
meson, $\pi^0$, $\eta$, or $\eta'$, due to their anomalous coupling to two photons.
The crucial input in the calculation of $a_\mu$ in the so-called pion-pole approximation \cite{Knecht:2001qf}
comes from the pseudoscalar transition form factor (TFF)
$F_{\pi^0\gamma^*\gamma^*}(Q_1^2,Q_2^2)$ (and analogously for $\eta$ and $\eta'$) at spacelike photon momenta.

$F(0,0)$ is determined by the known decay rates into two real photons. Experimental data for $F_{\pi^0}(Q^2,0)$
in the region $Q^2\lesssim 1$ GeV$^2$, where the bulk of the pion-pole contribution to $a_\mu$ arises \cite{Nyffeler:2016gnb},
have been obtained in particular by the CELLO collaboration \cite{Behrend:1990sr} 
and recently by BESIII \cite{Danilkin:2019mhd}.
In the calculation of $a_\mu$, a model of the double-virtual TFF is needed, for which data in the relevant
momentum region are still lacking.

In this note we shall review a set of holographic models for (large-$N_c$) quantum
chromodynamics (QCD) that have been studied previously \cite{Grigoryan:2007wn,Grigoryan:2008up,Grigoryan:2008cc,Hong:2009zw,Cappiello:2010uy} with respect to
their prediction of the pseudoscalar TFF 
and the resulting
prediction for the HLBL scattering contribution to $a_\mu$ in the pion-pole approximation \cite{Hong:2009zw,Cappiello:2010uy}. We compare in detail
the top-down model of Sakai and Sugimoto \cite{Sakai:2004cn,Sakai:2005yt}, which is only applicable 
at low momenta, and three bottom-up models, which have a simpler construction in the infrared but reproduce
qualitatively (and some also quantitatively) the short-distance constraints from perturbative QCD.
All these models are completely fixed, once $f_\pi$ and the mass of the $\rho$ meson are given.
Comparing with the experimental data reviewed in \cite{Danilkin:2019mhd}, which include preliminary low-energy data
for $F_{\pi^0}(Q^2,0)$ from BESIII, we find a surprisingly good agreement for the bottom-up models, whereas
the Sakai-Sugimoto model performs well only at the lowest values of $Q^2$. 
Compared to the interpolators
used in Ref.~\cite{Cappiello:2010uy} and the new one proposed by Danilkin, Redmer, and Vanderhaeghen (DRV) in \cite{Danilkin:2019mhd}, the holographic
models unanimously predict a milder dependence on the difference $Q_1^2-Q_2^2$ at fixed sum $Q_1^2+Q_2^2$,
which should be testable by future experiments at BESIII. In fact, the behavior 
of the $\pi^0$ TFF for generic virtualities found
in the dispersive approach of Ref.~\cite{Hoferichter:2018kwz} and
in the recent lattice extrapolation of Ref.~\cite{Gerardin:2019vio}
is closer to that of the holographic results than that of the DRV interpolator.

In contrast to Ref.~\cite{Cappiello:2010uy}, we have used directly the holographic TFF to numerically evaluate
the HLBL contribution to $a_\mu$ using the full 3-dimensional integral formulae of Ref.~\cite{Nyffeler:2016gnb},
without recourse to an interpolator that permits the simplification to the 2-dimensional integral representation of
Ref.~\cite{Knecht:2001qf}. Indeed, the (extended) D'Ambrosio-Isidori-Portoles (DIP) interpolator \cite{DAmbrosio:1997eof} used in  Ref.~\cite{Cappiello:2010uy}
to include one of the holographic predictions for the curvature parameters in the TFF 
does not permit to also fit the full UV behavior of the holographic models, missing in particular the Brodsky-Lepage constraint \cite{Lepage:1979zb,*Lepage:1980fj,*Brodsky:1981rp},
which is respected qualitatively in all bottom-up models. This turns out to lead to an over-estimation of the single-virtual
TFF and, as a consequence, gives a result for $a_\mu$ that is \st{about 10\%}\hl{somewhat} higher. 
With full numerical evaluation, the bottom-up-holographic results
span the range \hl{$a_\mu^{\pi^0}=6.1(4)\cdot 10^{-10}$}, which turns out to be well in line with the recent result obtained with
the DRV interpolator fitted to the world data for $\pi^0$ \cite{Danilkin:2019mhd} and also with the lattice results of Ref.~\cite{Gerardin:2019vio}.

\section{Holographic QCD models}

The AdS/CFT conjecture \cite{Maldacena:1997re} has led to a new approach to studying strongly 
interacting non-Abelian gauge theories in the limit
of large color number $N_c$. Originally formulated for superconformal field theories only, in particular
for the dual pair of type-IIB supergravity on AdS$_5\times S^5$ and four-dimensional
$\mathcal N=4$ super-Yang-Mills theory, it was quickly
realized that this can also be employed for studying nonconformal systems. Indeed, a most fruitful application
has been super-Yang-Mills theories at finite temperature, where temperature introduces a scale and also
breaks supersymmetry \cite{Witten:1998zw}. A similar procedure was proposed by Witten \cite{Witten:1998zw}
to model nonsupersymmetric low-energy Yang-Mills theory, namely
as a circle-compactified five-dimensional super-Yang-Mills theory whose dual is the near-horizon geometry of
D4 branes in type-IIA supergravity. Below the Kaluza-Klein mass scale introduced in compactifying a superfluous spatial
direction, only the Yang-Mills fields remain massless. Discarding the corresponding Kaluza-Klein modes (which
cannot be made arbitrarily heavy without leaving the supergravity approximation), one thus obtains 
a model for four-dimensional Yang-Mills theory given by type-IIA supergravity on a six-dimensional space
with the topology of a Euclidean black hole, i.e., a spacetime that is cut off smoothly when the radius of the compactifying
circle goes to zero. 

In Ref.~\cite{Sakai:2004cn,Sakai:2005yt} Sakai and Sugimoto succeeded in constructing a
holographic dual with chiral quarks and non-Abelian chiral symmetry breaking
by introducing $N_f$ probe D8 and anti-D8 branes localized in the extra dimension of the Witten model. 
The chiral symmetry $U(N_f)_\mathrm{L}\times  U(N_f)_\mathrm{R}$ of separated stacks of branes is broken spontaneously
to the diagonal subgroup by the necessity of connecting the D8-\AD\ pairs in the terminating
bulk geometry. The resulting model is a geometric realization of the so-called hidden local symmetry approach
to chiral symmetry breaking, correctly implementing the non-Abelian flavor anomalies through the Chern-Simons
term of the D8-branes.

Already before this fully top-down string-theoretic construction, bottom-up models which realize a confining geometry
by some cutoff of the bulk geometry have been created, where the bulk flavor gauge fields are introduced by hand
and a spontaneous
breaking of chiral symmetry is engineered either by an extra bifundamental scalar field or by
suitable boundary conditions. While the (Witten-)Sakai-Sugimoto (SS) model provides a most natural realization of
the infrared phenomena of confinement and chiral symmetry breaking, it lacks a viable ultraviolet completion
that would make contact to QCD at higher energies. Bottom-up models instead retain conformal symmetry in the ultraviolet and break
it forcibly in the infrared.

Following Ref.~\cite{Cappiello:2010uy} we shall focus on the most economical and thus also maximally predictive models which
use only the pion decay constant and the mass of the $\rho$ meson as input. In the hard-wall (HW) models, AdS$_5$ space
with metric
\be
ds^2=z^{-2}(\eta_{\mu\nu}dx^\mu dx^\nu - dz^2)
\ee
(and conformal boundary at $z=0$)
is simply cut off at some finite value of the radial coordinate $z_0$, whereas in the soft-wall (SW) models a nontrivial
dilaton background is used to produce a discrete spectrum with Regge behavior while having $z_0=\infty$.

Both the top-down and the various bottom-up models eventually describe bulk flavor gauge fields dual to vector and axial
vector mesons through a $U(N_f)\times U(N_f)$
Yang-Mills action in a curved five-dimensional background (with or without nontrivial dilaton background),
\be
S_{\rm YM} \propto \;\text{tr}\int d^4x \int_0^{z_0} dz\,e^{-\Phi(z)}\sqrt{-g}\, g^{PR}g^{QS}
\left(\mathcal{F}^\mathrm{L}_{PQ}\mathcal{F}^\mathrm{L}_{RS}
+\mathcal{F}^\mathrm{R}_{PQ}\mathcal{F}^\mathrm{R}_{RS}\right),
\ee
where $P,Q,R,S=0,\dots,3,z$ and $\mathcal{F}_{MN}=\partial_M \mathcal{B}_N-\partial_N \mathcal{B}_M-i[\mathcal{B}_M,\mathcal{B}_N]$.

In the SS model this action is obtained after truncating the nonpolynomial Dirac-Born-Infeld action
and integrating over the four-sphere wrapped by the D8 branes. 
There $g_{MN}$ is the induced metric on the D8 branes and 
$z_0$ is the value of the radial coordinate where the D8 and $\overline{\text{D8}}$ branes
connect, which is the extremal value of the geometry for antipodal branes (as we shall assume in what follows). 
The chiral field $U(x)=\exp[2i\pi(x)/f_\pi]$ appears as the holonomy
$U(x)=P\exp i\int \mathcal{B}$ integrated radially along the two connected branes
(or, in radial gauge, in the asymptotic behavior of the spatial components of $\mathcal{B}$).
Vector and axial vector fields are even and odd fields over the combined branes, which can be separated
into left and right contributions according to $\mathcal{F}^\mathrm{L,R}_{MN}=\mathcal{F}^\mathrm{V}_{MN}\mp \mathcal{F}^\mathrm{A}_{MN}$.
The physical vector and axial vector mesons correspond to the normalizable modes of these fields.
The D8 brane action also involves a Chern-Simons term which leads to the correct Wess-Zumino-Witten term \cite{Sakai:2004cn,Sakai:2005yt}
\be
S_{\rm CS}=\frac{N_c}{24\pi^2}\int\text{tr}\left(\mathcal{B}\mathcal{F}^2-\frac{i}2 \mathcal{B}^3\mathcal{F}
-\frac1{10}\mathcal{B}^5\right). 
\ee
(In the bottom-up models, where $\mathcal{B}^\mathrm{L}$ and $\mathcal{B}^\mathrm{R}$ fields appear separately, 
this is added by hand as $S_{\rm CS}^\mathrm{L}-S_{\rm CS}^\mathrm{R}$.)
The electromagnetic gauge field can be introduced as a non-dynamical
background field through a nonzero boundary value for the vector gauge field
with generator equal to the electric charge matrix, which naturally leads to vector meson dominance (VMD) \cite{Sakai:2005yt}.

In the following we collect some relevant results for the various models which will then be used to
determine the pseudoscalar-photon transition form factors. For more details see Ref.~\cite{Cappiello:2010uy} and
references therein.

\subsection{Sakai-Sugimoto model}

Using a dimensionless coordinate $Z$ that runs from $-\infty$ to $+\infty$ along the connected D8-\AD\ branes,
the Yang-Mills part of the action of the SS model reads \cite{Sakai:2004cn,Sakai:2005yt}
\be\label{SD8F2}
S_{\rm YM}=-\kappa\,\Tr\int d^4x \int_{-\infty}^\infty dZ\left[
\frac12 (1+Z^2)^{-1/3}\eta^{\mu\rho}\eta^{\nu\sigma}\mathcal{F}_{\mu\nu}\mathcal{F}_{\rho\sigma}
+(1+Z^2)\MKK^2\eta^{\mu\nu}\mathcal{F}_{\mu Z}\mathcal{F}_{\nu Z}\right]
\ee
with $\kappa={\lambda N_c}/(216\pi^3)$ and $\lambda=g_{\rm YM}^2 N_c$.

Massive vector and axial vector mesons arise as even and odd eigenmodes
of $\mathcal{B}_\mu^{(n)}=\psi_n(Z) v^{(n)}_\mu(x)$ with eigenvalue equation
\be\label{psin}
-(1+Z^2)^{1/3}\6_Z\left[ (1+Z^2)\6_Z \psi_n \right]=\lambda_n \psi_n,\quad \psi_n(\pm\infty)=0.
\ee
The lowest mode $v_\mu^{(1)}$ is 
interpreted as the isotriplet $\rho$ meson (or the $\omega$ meson
for the U(1) generator) with mass $m_\rho^2=\lambda_1 \MKK^2$.
The numerical result $\lambda_1=0.669314\ldots$
fixes the Kaluza-Klein mass of the SS model (the inverse radius of the $x^4$ circle where
the D8 and \AD\ branes are localized antipodally) to $\MKK=949$~MeV.

The holographic factor of the pion wave function is associated 
to the derivative of the (non-normalizable) zero-mode of (\ref{psin}) of the axial vector sector,
\be
\alpha^{\rm SS}(Z)=\frac{\pi}2 \arctan(Z).
\ee
The latter appears in the large gauge transformation that would be needed to enforce
a radial gauge $\mathcal{B}_Z=0$, which relocates the pion field from the holonomy 
$U(x)=P\exp i\int_{-\infty}^\infty dZ\mathcal{B}_Z$ to nontrivial boundary conditions on $\mathcal{B}_\mu$
\cite{Sakai:2004cn}.
Either way,
the pion decay constant turns out to be given by $f_\pi^2=\lambda N_c\MKK^2/(54\pi^4)$.
Choosing $f_\pi=92.4$ MeV corresponds to $\kappa=0.00745$ or $\lambda\approx 16.63$ for $N_c=3$.

A background photon field $A_\mu(x)$ is included by setting $\psi(\pm\infty)=1$ for 
$\mathcal{B}_\mu=e\mathcal{Q}A_\mu(x)\psi(Z)$ with $\mathcal{Q}={\rm diag}(\frac23,-\frac13,-\frac13)$. 
A real photon with $q^2=0$
corresponds to the trivial solution $\psi(Z)\equiv 1$, whereas a virtual photon with spacelike
momentum $Q^2>0$ is described by solutions where $\lambda_n\to -Q^2/\MKK^2$.
This defines the so-called bulk-to-boundary propagator, which will be denoted by $\mathcal{J}$ in the following,
\be
(1+Z^2)^{1/3}\6_Z\left[ (1+Z^2)\6_Z \mathcal{J} \right]=\frac{Q^2}{\MKK^2}\mathcal{J},\quad \mathcal{J}(Q,Z=\pm\infty)=1.
\ee

While the original SS model is a strictly chiral model (and we shall stick to that), masses for quarks can be included
through worldsheet instantons \cite{Aharony:2008an,Hashimoto:2008sr} which leads to correct Gell-Mann--Oakes--Renner relations.
Moreover, at order $1/N_c$ the axial U(1)$_A$ is broken in the SS model, which thereby 
includes a Witten-Veneziano mechanism \cite{Witten:1979vv,Veneziano:1979ec} for giving mass to
the $\eta_0$ pseudoscalar according to \cite{Sakai:2004cn} $m_{0}^2=N_f\lambda^2\MKK^2/(27\pi^2 N_c)$, which
is in the right ballpark to account for realistic pseudoscalar meson masses \cite{Brunner:2015oga}.

However, as mentioned above, the SS model is not asymptotically AdS, but has a diverging dilaton in the UV.
It therefore can serve as a dual to QCD only at small momenta.

\subsection{Bottom-up models}

\subsubsection{Hard-wall model with bi-fundamental scalar (HW1)}

In the hard-wall model of Ref.~\cite{Erlich:2005qh,DaRold:2005mxj}, a bi-fundamental bulk scalar $X$
is introduced, with a five-dimensional mass term determined by the scaling dimension $\Delta=3$ of the chiral-symmetry
breaking order parameter $\bar q q$ of the boundary theory. At a finite value $z_0$, a cutoff of AdS$_5$ space is
imposed with boundary conditions $\mathcal{F}^\mathrm{L,R}_{z\mu}=0$.

At zero 4-momentum, the modulus of the scalar field has the form $v(z)=m_q z+\sigma z^3$, where $m_q$ is interpreted as
the explicit symmetry-breaking quark mass of the boundary theory, and $\sigma$ the $\bar q q$ condensate.
The pion field comes from both the phase of $X$ and the longitudinal components of the axial gauge fields.
In the chiral limit, its holographic wave function can be given in closed form as \cite{Grigoryan:2007wn,Grigoryan:2008up}
\be
\Psi(z)=\Gamma({\textstyle{\frac23}})\left(\xi z^3/2\right)^{1/3}
\left[I_{-1/3}(\xi z^3)-\frac{I_{2/3}(\xi z_0^3)}{I_{-2/3}(\xi z_0^3)}I_{1/3}(\xi z^3) \right],
\ee
where $\xi=g_5\sigma/3$ and $g_5^2=12\pi^2/N_c$.
The parameter $\xi$ is fixed through the relation
\be\label{HW1sigma}
\frac{6\pi^2}{N_c}f_\pi^2=\frac{\Gamma(\frac23)}{\Gamma(\frac43)}\frac{I_{2/3}(\xi z_0^3)}{I_{-2/3}(\xi z_0^3)}(\xi/2)^{2/3},
\ee
which yields $\xi=(0.424 \,\text{GeV})^3$ for $f_\pi=92.4$ MeV.

Vector mesons have a holographic wave function given by
\be\label{psinHW}
\partial_z\left[\frac1z \partial_z \psi_n(z)\right]+\frac1z M_n^2 \psi_n(z)=0
\ee
with boundary conditions $\psi_n(0)=\psi'_n(z_0)=0$, solved by $\psi_n(z)\propto zJ_1(M_n z)$ with $M_n$ determined by
the zeros of the Bessel function $J_0$. Thereby $z_0$ gets fixed to $z_0=\gamma_{0,1}/m_\rho$, where $\gamma_{0,1}= 2.40483\ldots$
gives the first zero of $J_0$. For $m_\rho=775$ MeV, one obtains $z_0=3.103 \, \text{GeV}^{-1}$.

The vector bulk-to-boundary propagator is obtained by replacing $M_n^2\to -Q^2$ and the boundary conditions by
$\mathcal{J}(Q,0)=1$ and $\partial_z \mathcal{J}(Q,z_0)=0$, which gives
\be\label{HWVF}
\mathcal{J}(Q,z)=
Qz \left[ K_1(Qz)+\frac{K_0(Q z_0)}{I_0(Q z_0)}I_1(Q z) \right].
\ee

\subsubsection{Hirn-Sanz model (HW2)}

The hard-wall model by Hirn and Sanz \cite{Hirn:2005nr} (called HW2 in \cite{Cappiello:2010uy}) 
refrains from introducing a matrix-valued scalar field for the purpose of chiral symmetry breaking.
This is instead implemented in a way that is very similar to the SS model, but in a much simpler manner.
The pion field is also built from Wilson lines running along the holographic direction, $U(x)=\xi_\mathrm{R}(x)\xi_\mathrm{L}(x)$
with $\xi_\mathrm{L,R}=P\exp(- i\int_0^{z_0}dz \mathcal{B}_z^\mathrm{L,R})$.
Vector and axial vector mesons are distinguished by different boundary conditions on the hard wall,
Neumann for vector and Dirichlet for axial vector mesons, which is precisely what distinguishes
them in the SS model at the point where D8 and \AD\ branes connect.
As in the SS model, the pion wave function appears as the derivative of a non-normalizable zero mode
$\alpha$ of the axial vector field,
\be
\alpha^{\rm HW2}(z)=1-\frac{z^2}{z_0^2}.
\ee
The vector meson field equation is the same as in the HW1 model, and therefore also the vector
bulk-to-boundary propagator $\mathcal{J}$, Eq.~(\ref{HWVF}), as well as the value of $z_0$.

\subsubsection{Soft-wall model}

Soft-wall models were originally introduced to achieve a Regge-type spectrum of mesons \cite{Karch:2006pv,Grigoryan:2007my}.
Prescribing by hand a nontrivial background dilaton field $\Phi(z)=\kappa^2 z^2$ leads to
$M_n^2=4\kappa^2(n+1)$ for the vector mesons, where $\kappa=m_\rho/2$. Since now $z_0=\infty$, the boundary conditions
in the IR are replaced by the requirement of normalizability.
 
The vector field equation is given by (\ref{psinHW}) with $1/z$ replaced by $e^{-\kappa^2 z^2}/z$. A closed
form solution for $\mathcal{J}$ can be given in terms the confluent hypergeometric function of the second kind $U(a,b,z)$:
\be
\mathcal{J}^{\rm SW}(Q,\kappa,z)= \frac{U(\textstyle{\frac{Q^2}{4\kappa^2}},0,(\kappa z)^2)}{U(\textstyle{\frac{Q^2}{4\kappa^2}},0,0)}
=\Gamma(1+\textstyle{\frac{Q^2}{4\kappa^2}})U(\textstyle{\frac{Q^2}{4\kappa^2}},0,(\kappa z)^2).
\ee

In this model chiral symmetry breaking is implemented in a less clear manner. The soft-wall model considered
in \cite{Cappiello:2010uy} (and also here) follows Ref.~\cite{Grigoryan:2008up}, where the pion wave function is
assumed to be Gaussian,
\be
\alpha^{\rm SW}(z)=e^{-\kappa^2 z^2}.
\ee

\section{Pseudoscalar-photon transition form factors}

The pseudoscalar-photon TFF is defined by
\be
\int d^4x e^{-i q_1\cdot x}\langle P(q_1+q_2)|T\left\{ J_\mu^{e.m.}(x) J_\nu^{e.m.}(0)\right\}|0\rangle
=\epsilon_{\mu\nu\rho\sigma}q_1^\rho q_2^\sigma F_{P\gamma^*\gamma^*}(Q_1^2,Q_2^2),
\ee
where $Q_{1,2}^2=-q_{1,2}^2$. 
For $P=\pi^0$
the value for real photons is determined by the axial anomaly according to
\be\label{Fpi00}
F_{\pi^0\gamma^*\gamma^*}(0,0)=\frac{N_c}{12\pi^2 f_\pi}.
\ee

\subsection{Holographic predictions}

In the following we shall denote the normalized TFF by $K$,
\be
K(Q_1^2,Q_2^2)\equiv F(Q_1^2,Q_2^2)/F(0,0),
\ee
and consider the various holographic predictions in the chiral limit. 
The derivation of the holographic result has been discussed in detail in Refs.~\cite{Grigoryan:2008up,Grigoryan:2008cc,Cappiello:2010uy,Stoffers:2011xe}
for the various models, 
in Ref.~\cite{Hong:2009zw} also by including the effects of finite quark masses in the HW1 model.
It is determined by the universal form of the Chern-Simons action needed to take into account
the chiral anomaly, which leads to an integral over two bulk-to-boundary propagators for the virtual
photons times the holographic pion wave function,
\be
K(Q_1^2,Q_2^2)=-\int_0^{z_0} \mathcal{J}(Q_1,z) \mathcal{J}(Q_2,z) \partial_z \alpha(z) dz.
\ee
In the case of the HW1 model, this needs to be corrected by a boundary term in the infrared \cite{Grigoryan:2008up}, because
the HW1 pion wave function $\Psi(z_0)\not=0$, 
\be
K^{\rm HW1}(Q_1^2,Q_2^2)=-\int_0^{z_0} \mathcal{J}(Q_1,z) \mathcal{J}(Q_2,z) \partial_z \Psi(z) dz
+\mathcal{J}(Q_1,z_0) \mathcal{J}(Q_2,z_0) \Psi(z_0).
\ee

\subsection{Low-$Q^2$ behavior and comparison to data with $Q_2^2=0$}

The behavior of $K(Q_1^2,Q_2^2)$ at virtualities below 1 GeV$^2$ are decisive for the HLBL contribution to $a_\mu$.
It is therefore of interest to parametrize the low-$Q^2$ behavior by the first few Taylor coefficients,
which following \cite{Cappiello:2010uy} we define by
\be\label{Kabc}
K(Q_1^2,Q_2^2)=1+\hat\alpha(Q_1^2+Q_2^2)+\hat\beta Q_1^2 Q_2^2+\hat\gamma(Q_1^4+Q_2^4)+O(Q^6).
\ee
The slope parameter $\hat\alpha$ is often quoted as $a_\pi=-\hat\alpha m_\pi^2$ or $\Lambda^2=-1/\hat\alpha$.

\begin{table}
\bigskip
\begin{tabular}{l|c|c|c|c|c|c|}
\toprule
 Model  & $\Lambda^2[\text{GeV}^2]$ & $\hat\alpha[\text{GeV}^{-2}]$  & $\hat\beta[\text{GeV}^{-4}]$  & $\hat\gamma[\text{GeV}^{-4}]$  & $\bar F(0,\infty)$ & $\bar F(\infty,\infty)$ 
 \\
\colrule
SS & 0.489 & -2.043 &  4.56 & 3.55 & 0 & 0 \\
\colrule
HW1 & 0.627 & -1.595 &  3.01 & 2.63 & 1.00 & 1.00  \\ 
HW2 & 0.554 & -1.805 &  3.65 & 3.06 & 0.62 & 0.62 \\
SW & 0.601 & -1.665 &  3.56 & 2.76 & 0.89 & 0.89 \\
\colrule
DIP1 & 0.568 & -1.760 & 3.33 & 3.78 & $\infty$ & 1.00 \\
DIP2 & 0.568 & -1.760 & 3.33 & 3.88 & $\infty$ & 1.00 \\
\colrule
DRV4 & 0.574 & -1.742 & $\infty$ & 3.04 & 0.85 & 0.85 \\
DRV9 & 0.611 & -1.637 & $\infty$ & 2.68 & 0.90 & 0.90 \\
\botrule
\end{tabular}
\caption{IR and UV behavior of $K(Q_1^2,Q_2^2)$ in the top-down holographic SS model and the three bottom-up models HW1, HW2, and SW considered here (all in the chiral limit), with IR coefficients as defined in (\ref{Kabc})
and $\Lambda^2\equiv-1/\hat\alpha$; the UV behavior is described by $\bar F=F/F^\infty$, where $F^\infty$ is the LO QCD result (\ref{Finfinity}). Also given are the corresponding quantities of the interpolators (\ref{DIP1}) and (\ref{DIP2}) (DIP1 and DIP2) used in
\cite{Cappiello:2010uy} to approximate the bottom-up models with respect to $\hat\alpha$ and $\hat\beta$, and the DRV interpolator (\ref{DRV}) used in \cite{Danilkin:2019mhd} to fit the currently available data up to 4 or 9 GeV$^2$.}
\label{IRUVbehavior}
\end{table}

In Table \ref{IRUVbehavior}, these parameters are given for the various holographic models.
Until recently, the experimental world average for $\hat\alpha$, which was dominated by the result of the CELLO collaboration
\cite{Behrend:1990sr}, read $\hat\alpha=-1.76(22)\,\text{GeV}^{-2}$. In Ref.~\cite{Cappiello:2010uy} this
was taken as an indication that the SS model, which has $\hat\alpha^{\rm SS}=-2.043$, should be discarded
as a viable model for the pion TFF. A recent analysis of Dalitz decays of $\pi^0$ from NA62 gave \cite{TheNA62:2016fhr}
$\hat\alpha=-2.02(31)\,\text{GeV}^{-2}$, leading to the new world average \PDG\ $-1.84(17)\,\text{GeV}^{-2}$.
The previous world average and its error had covered all results of the bottom-up models, while being at some tension
with the SS model, but now also the HW1 model is slightly disfavored, while optimal agreement is found for the HW2 model.

So far, experimental data do not allow to discriminate between values for the curvature parameter $\hat\gamma$, and even less so
for the double-virtual curvature parameter $\hat\beta$. 
(As discussed in more detail below, Ref.~\cite{Cappiello:2010uy} introduced the interpolators DIP1 and DIP2, defined in (\ref{DIP1}) and (\ref{DIP2}), 
such that they reproduce the average of the bottom-up holographic results for
$\hat\alpha$ and $\hat\beta$. As shown in Table \ref{IRUVbehavior} they both overestimate the holographic results for $\hat\gamma$.)

\begin{figure}
\bigskip
\includegraphics[width=0.7\textwidth]{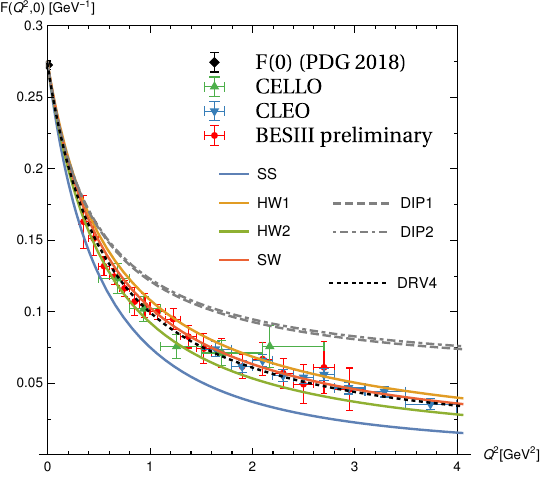}
\caption{Data for the $\pi^0$ TTF from CELLO, CLEO, BESIII-preliminary as compiled in Fig.~3 of Ref.~\cite{Danilkin:2019mhd} compared to the results of the various holographic models, the interpolators (DIP1, DIP2) proposed in Ref.~\cite{Cappiello:2010uy} with curvature parameter $\hat\beta$ matched to the bottom-up holographic models, and the recent improved interpolator of Ref.~\cite{Danilkin:2019mhd} fitting the experimental data up to 4 GeV (DRV4). (The fit up to 9 GeV used in the evaluation of $a_\mu$ happens to agree so closely with the SW prediction
that it could not be distinguished within the line thickness of this plot.)
The result of the recent dispersion relation study of Ref.~\cite{Hoferichter:2018kwz}
(not shown to avoid overcrowding) 
lies right in between the SW (red) and HW1 (orange) result, with the lower end of the error band given in Ref.~\cite{Hoferichter:2018kwz}
nearly coinciding with the SW result.
}
\label{fig3comp}
\end{figure}

In Fig.~\ref{fig3comp}, the holographic results are compared with spacelike $\pi^0$ TFF data at $Q^2\le 4\,\text{GeV}^2$
as compiled in Fig.~3 of Ref.~\cite{Danilkin:2019mhd}, which includes preliminary data from BESIII down to 0.3 GeV$^2$.
At the lowest available $Q^2$ value, all holographic results are within the experimental error bar, while at higher $Q^2$
the SS model underestimates the experimental result. All the bottom-up holographic results are however found to agree
remarkably well with the data. As we shall discuss in the next
subsection, this correlates with their behavior at very high $Q^2$. (On the other hand, the DIP interpolators, which aim
to have correct behavior at high $Q_1^2=Q_2^2$ while matching part of the low-$Q^2$ behavior of
the bottom-up holographic models, are shown to provide a poor fit to the single-virtual TFF data.)

Also shown in Fig.~\ref{fig3comp} is the fit with a new interpolator of Ref.~\cite{Danilkin:2019mhd} 
taking into account the experimental data up to 4 GeV (DRV4); the fit up to 9 GeV that was
used in the evaluation of $a_\mu$ therein happens to coincide with the SW prediction within the line thickness.
The result of the recent dispersion relation study of Ref.~\cite{Hoferichter:2018kwz}
is slightly higher and in between the SW and the HW1 prediction, with the lower end of the error
band obtained in Ref.~\cite{Hoferichter:2018kwz} nearly coinciding with the SW result
(not shown in Fig.~\ref{fig3comp} to avoid overcrowding).

\begin{figure}
\bigskip
\includegraphics[width=0.7\textwidth]{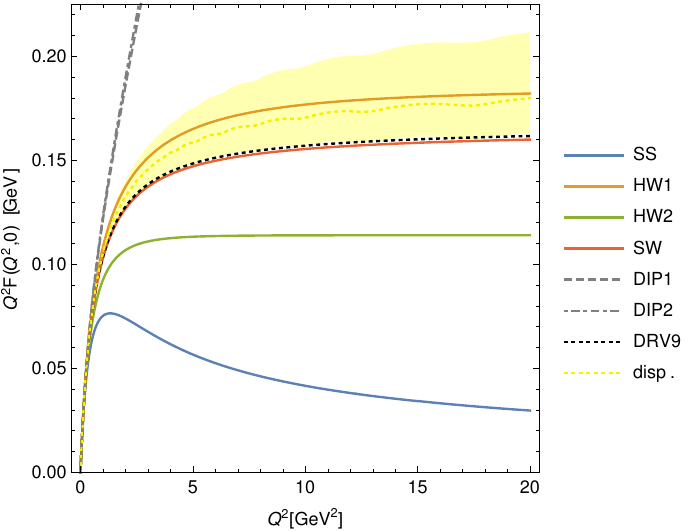}
\caption{Comparison of the 
results of the various holographic models and the interpolators (DIP1, DIP2) proposed in Ref.~\cite{Cappiello:2010uy}
with the interpolator of Ref.~\cite{Danilkin:2019mhd} fitting all experimental data up to 9 GeV (DRV9)
and the 
result of the dispersion relation study of Ref.~\cite{Hoferichter:2018kwz}
(yellow band for the estimated error with dotted points for the central result). 
}
\label{figsingvirtlarge}
\end{figure}

In Fig.~\ref{figsingvirtlarge} the various single-virtual pion TFF results are shown together with the result of the dispersion
relation study of Ref.~\cite{Hoferichter:2018kwz} for higher $Q^2$.
Here the experimental data (not shown) scatter more strongly, but are covered well
by the error band given in Ref.~\cite{Hoferichter:2018kwz}. At these higher values of $Q^2$
the central result of the dispersive approach is still in between the SW and HW1 results, but
somewhat closer to the latter; the almost coincident SW and DRV9 lines are just below the error band
of the dispersive result. The HW2 model has similar asymptotic behavior, but is quantitatively below
the dispersive result, whereas the pion TFF of the SS model evidently decays faster than $1/Q^2$ at large $Q^2$.

\subsection{Short-distance behavior and the double virtual case}

Perturbative QCD (pQCD)
predicts a factorization of the double virtual TFF into a perturbatively calculable hard-scattering kernel and
a nonperturbative meson distribution amplitude \cite{Lepage:1979zb,*Lepage:1980fj,*Brodsky:1981rp,Efremov:1979qk}. To leading order (LO), an
asymptotic pion distribution function $\Phi_\pi(x)=6x(1-x)$, where $x$ and $(1-x)$ are the momentum fractions
of a collinear quark/anti-quark state,
yields \cite{Danilkin:2019mhd}
\be
K^\infty(Q_1^2,Q_2^2)=\frac{8\pi^2 f_\pi^2}{Q_1^2+Q_2^2}f(w)
\ee
with $w=(Q_1^2-Q_2^2)/(Q_1^2+Q_2^2)$ and
\be\label{fw}
f(w)=\frac1{w^2}-\frac{1-w^2}{2w^3}\ln\frac{1+w}{1-w},
\ee
which equals unity at $w=\pm 1$ and drops to $f(0)=2/3$ at the symmetric point. This corresponds to
asymptotic behavior
\be\label{Finfinity}
F^\infty(Q^2,0)=\frac{2 f_\pi}{Q^2},\qquad
F^\infty(Q^2,Q^2)=\frac{2 f_\pi}{3 Q^2}.
\ee
Perturbative corrections have been worked out in Ref.~\cite{Melic:2002ij} to order $\alpha_s$ and $\alpha_s^2$; they lead to a moderate
reduction of the LO result for $Q^2 \gtrsim 2\, \text{GeV}^2$, but evidently much higher values of $Q^2$ are needed for the experimental data to approach the perturbative regime.

As shown in \cite{Grigoryan:2008up}, the HW1 model has an asymptotic limit equal to the full LO pQCD result, when the parameter $\sigma$
is fixed according to (\ref{HW1sigma}). Amazingly, exactly the same $w$-dependence arises for $Q^2\to\infty$:
\bea
K^{\rm HW1}(Q_1^2,Q_2^2)&\to& \frac{8\pi^2 f_\pi^2}{Q_1^2+Q_2^2}\sqrt{1-w^2}
\int_0^\infty d\xi\, \xi^3 K_1(\xi\sqrt{1+w}) K_1(\xi\sqrt{1-w})
\nonumber\\
&&=\frac{8\pi^2 f_\pi^2}{Q_1^2+Q_2^2}f(w),
\eea
where $\xi=Qz$ and $f(w)$ as given in (\ref{fw}).

Apart from a different overall factor, the same form is found in the other bottom-up models, which is
a direct consequence of the asymptotic AdS geometry \cite{Cappiello:2010uy}. 

In the HW2 model,
the factor $8\pi^2 f_\pi^2$ is replaced by $4/z_0^2$. With $z_0=3.103\, \text{GeV}^{-1}$ in order to
reproduce the value of the $\rho$ meson mass, this corresponds to $\approx 61.6\%$ of the LO pQCD result.\footnote{As
noted in Ref.~\cite{Cappiello:2010uy} (without doing so), 
an asymptotic limit equal to that of the HW1 model and thus the full LO pQCD result
could be achieved by choosing instead $z_0=1/(\sqrt{2}\pi f_\pi)$, however at the cost of a $\rho$ meson mass of 987 MeV.
This would spoil the agreement of the HW2 model with the low-energy data shown in Fig.~\ref{fig3comp};
it would then lie above all of the experimental error bars therein for virtualities up to $Q^2=2 \text{GeV}^2$.
(Note that in the original paper of Hirn and Sanz \cite{Hirn:2005nr}, $f_\pi$, $m_\rho$, and also the asymptotic limit are fitted,
resulting in $N_c\approx 4.3$, thus being no option here, as it would give a wrong prefactor for the Chern-Simons action.)}

In the SW model, the prefactor is instead $4\kappa^2$, with the same integral resulting in the limit $Q^2\gg \kappa^2$
because
\be
\mathcal{J}^{\rm SW}(Q,\kappa,z)= \frac{U(\textstyle{\frac{Q^2}{4\kappa^2}},0,(\kappa z)^2)}{U(\textstyle{\frac{Q^2}{4\kappa^2}},0,0)}
\to (Qz)\, K_1(Qz)
\ee
and the extra factor $e^{-\kappa^2\xi^2/Q^2}$ in the pion wave function becoming negligible. With $\kappa=m_\rho/2$, the overall
factor is thus reduced to $m_\rho^2/(8\pi^2 f_\pi^2) \approx 0.893$.

The top-down holographic QCD model of Sakai and Sugimoto on the other hand is only meaningful in the low-energy limit.
Here the asymptotic geometry is not AdS$_5$ but instead six-dimensional with a diverging dilaton.
Considering nevertheless the limit $Q/\MKK\to\infty$, we find
\be
\mathcal{J}^{\rm SS}(Q,Z) \to \left(1+3\bar Q \zeta^\frac13\right) e^{-3 \bar Q \zeta^\frac13}
\ee
with $\bar Q=Q/\MKK$, $\zeta=\arctan(1/Z)$,
and thus
\be
K^{\rm SS}(Q_1^2,Q_2^2)\to
\frac{16}{9\pi}\left(\frac{2\MKK^2}{Q_1^2+Q_2^2}\right)^\frac{3}{2}
\frac{2+5\sqrt{1-w^2}}{(\sqrt{1-w}+\sqrt{1+w})^5}.
\ee
This different $w$-dependence turns out to be rather similar quantitatively to that of the bottom-up models:
At constant $Q_1^2+Q_2^2$, the ratio of symmetrically double-virtual $K(Q^2/2,Q^2/2)$ over single-virtual $K(Q^2,0)$, which
at LO pQCD equals 2/3, asymptotes to $7/(8\sqrt{2})\approx 0.62$ in the SS model. 
The more significant 
difference is in the faster falloff $\sim Q^{-3}$,
which highlights the fact that the SS model is applicable only in the low-$Q^2$ regime.

In Table \ref{IRUVbehavior} the short-distance behavior of the various models is listed in comparison to the LO pQCD result, defining
$\bar F(Q_1^2,Q_2^2)=F(Q_1^2,Q_2^2)/F^\infty(Q_1^2,Q_2^2)$. The DIP interpolators (defined below in (\ref{DIP1}) and (\ref{DIP2})) reproduce by construction
the correct limit for $Q_1^2=Q_2^2\to\infty$, but cannot satisfy the Brodsky-Lepage constraint \cite{Lepage:1979zb,*Lepage:1980fj,*Brodsky:1981rp} of nontrivial asymptotic $Q^{-2}$ behavior
in the single virtual case. The DRV interpolator 
that was recently proposed in \cite{Danilkin:2019mhd} is defined by
\bea\label{DRV}
K^{\rm DRV}(Q_1^2,Q_2^2)&=&\frac1{1+(Q_1^2+Q_2^2)/\Lambda^2}f(w_\Lambda),\nonumber\\
w_\Lambda&\equiv&\left( \frac{(Q_1^2-Q_2^2)^2+\Lambda^4}{(Q_1^2+Q_2^2)^2+\Lambda^4} \right)^{1/2}.
\eea
It
has by construction an asymptotic $Q^{-2} f(w)$ behavior, without
being fixed to the LO pQCD prefactor. The overall factor instead depends on the value of the single fitting parameter $\Lambda$
and is given in Table \ref{IRUVbehavior} for the two fits of the pion TFF data presented in \cite{Danilkin:2019mhd}.
Notice that the DRV interpolator has a logarithmically divergent curvature in the double-virtual case as $Q_{1,2}^2\to0$.

\begin{figure}
\bigskip
\includegraphics[width=0.46\textwidth]{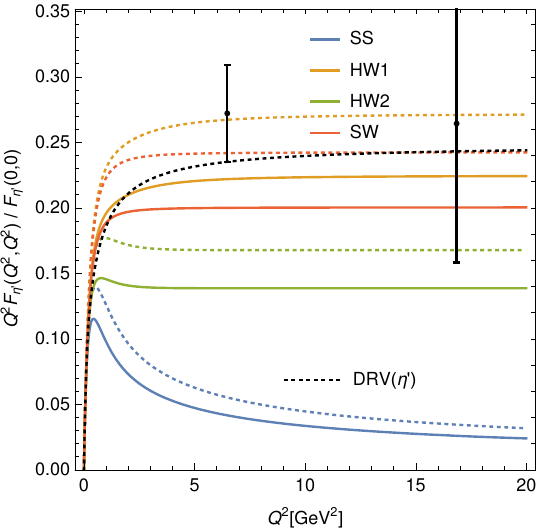}\quad\includegraphics[width=0.45\textwidth]{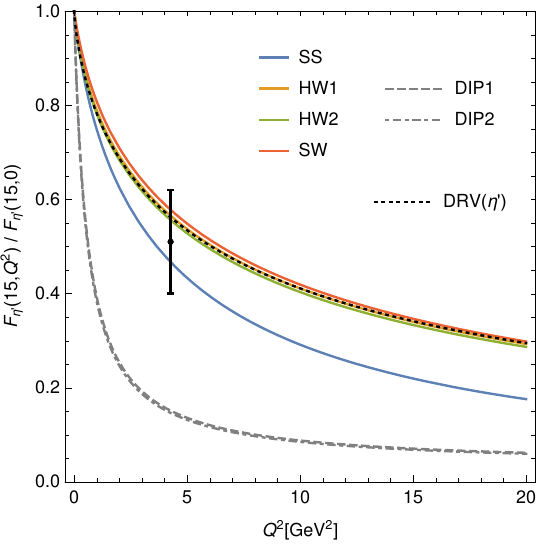}
\caption{Comparison of the results for symmetric and asymmetric double-virtual TFF with experimental data for $\eta'$ from BaBar \cite{BaBar:2018zpn}
and with DRV($\eta'$) from Ref.~\cite{Danilkin:2019mhd} which is determined by a fit to experimental data for $F_{\eta'}(Q^2,0)$. 
Full lines correspond to the results of the various holographic models in the chiral limit; 
dash-dotted gray lines represent the DIP extrapolators of Ref.~\cite{Cappiello:2010uy} (left out in
the symmetric case where it almost exactly coincides with the HW1 result).
In the left plot, dotted lines correspond to an extrapolation where the mass scale within $K(Q_1^2,Q_2^2)$ has been scaled up
by 10\% to match the rescaling in the corresponding function of the DRV interpolator between DRV9 for pions to DRV($\eta'$);
in the right plot these are left out because here the changes are almost invisible.
}
\label{fig18comp}
\end{figure}

In \cite{Danilkin:2019mhd} the DRV interpolator has been compared with recent experimental data from
BaBar \cite{BaBar:2018zpn} for double virtual $F_{\eta'}(Q_1^2,Q_2^2)$. In Fig.~\ref{fig18comp} 
this comparison is extended to
the holographic models and the DIP interpolators for the normalized function $F_{\eta'}(Q_1^2,Q_2^2)/F_{\eta'}(0,0)$.
The bottom-up models with the largest UV prefactor, the HW1 and SW models, compare favorably with the data
for symmetric double virtual TFF, while 
the HW2 is further away, and the SS model clearly completely off. However, it should be kept in mind that for the application
to $a_\mu$ only comparatively small values of $Q^2$ will be relevant. In Fig.~\ref{fig18comp} also the asymmetric double virtual TFF
for $\eta'$ is considered, with one data point from BaBar \cite{BaBar:2018zpn}. Interestingly enough, the SS result
for the ratio $F_{\eta'}(15\text{GeV}^2,Q_2^2)/F_{\eta'}(15\text{GeV}^2,0)$ is not too far from the pQCD result, which at these momenta are closely
reproduced by the bottom-up models and the DRV interpolator, and all agree with the experimental value within errors, whereas the DIP ansatz fails to reproduce it.
Note, however, that the holographic QCD results refer to the chiral limit, so that this comparison is only
meaningful to the extent that the quark mass dependence of the function $K(Q_1^2,Q_2^2)$ is weak.\footnote{Fig.~\ref{fig18comp} also 
shows that an upscaling of the mass scale within $K(Q_1^2,Q_2^2)$ by 10\%, which corresponds approximately to the ratio
of the monopole fit parameter $\Lambda$ obtained in \cite{Danilkin:2019mhd} for $\eta'$ over that for $\pi^0$, brings the holographic results into better agreement
with the experimental results for the symmetric case while retaining the already good agreement with the one asymmetric data point.
Doing so produces actually a better agreement with the BaBar data when using the HW1 and SW results 
as a phenomenological interpolator than is achieved by the DRV interpolator.}

\begin{figure}
\includegraphics[width=0.7545\textwidth]{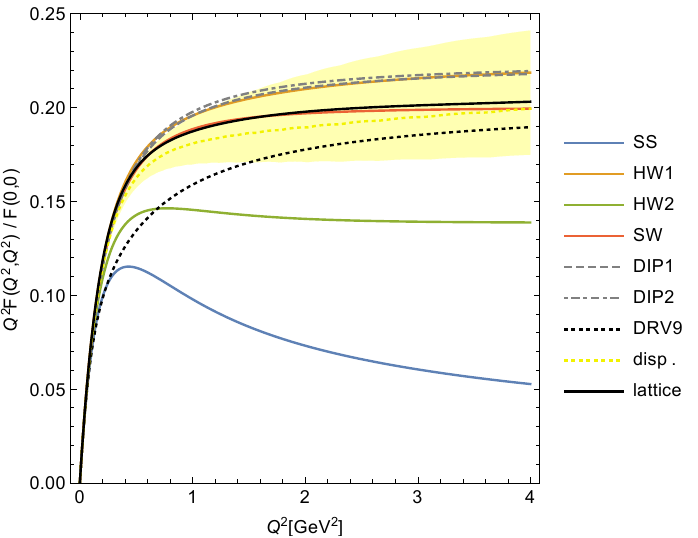}

\bigskip
\includegraphics[width=0.7545\textwidth]{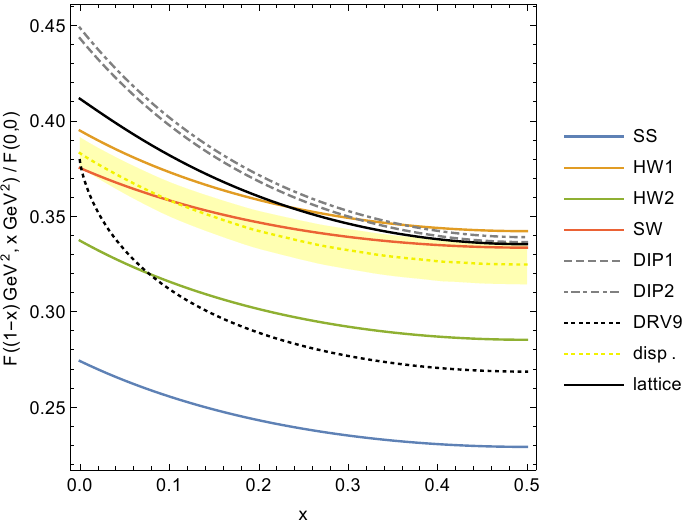}
\caption{Comparison of the predictions for the symmetrically double-virtual pion TFF $Q^2 F(Q^2,Q^2)/F(0,0)$ (upper plot)
and for the dependence on $x$ in the asymmetric $F((1-x)\text{GeV}^2,x\,\text{GeV}^2)/F(0,0)$, i.e., at $Q_1^2+Q_2^2=1\,\text{GeV}^2$,
(lower plot) including
the recent lattice extrapolations of Ref.~\cite{Gerardin:2019vio} (black line, corresponding to the central line of Fig.~7 therein)
and the dispersion theory results of Ref.~\cite{Hoferichter:2018kwz} (yellow band, with central result dotted).
}
\label{figKw}
\end{figure}

Unfortunately, there are not yet experimental data for the double-virtual pion TFF which would be particularly interesting at low values of $Q^2$, 
although these could be provided by BESIII around and below 1 GeV$^2$ in the future \cite{Danilkin:2019mhd}.
However, 
there exists a new lattice QCD calculation \cite{Gerardin:2019vio}, which extrapolates its results to arbitrary $Q_1^2$ and $Q_2^2$
with estimated errors that (at least when both $Q_1^2,Q_2^2>0$) are smaller than those of the dispersion theory results for the pion TFF of Ref.~\cite{Hoferichter:2018kwz}.
Both are compared with the holographic results and the DIP1,2 and DRV4 interpolators in Fig.~\ref{figKw}. In the symmetrically double-virtual case, the lattice result
for $Q^2 F(Q^2,Q^2)/F(0,0)$
is slightly above the central result of the dispersive approach, with its estimated error displayed again as a yellow band.
The predictions of the HW1 model and the SW model lie within this error band, with the SW model almost coinciding with the lattice result,\footnote{The near
coincidence of the HW1 model with the DIP results is due to the fact that the HW1 model asymptotes to the complete LO pQCD result,
whose value at the symmetric point has been used as an input for the DIP interpolator. The SW model asymptotes to 89\% of the LO pQCD result
(s.~Table \ref{IRUVbehavior}) which could be fortuitously the right amount at moderate energies.}
whereas the DRV interpolator is below the dispersive error band for $Q^2$ below 1.5 GeV$^2$.
The HW2 model is closer to the dispersive results below 0.7 GeV$^2$, but significantly further away at higher virtualities. 
The SS model result drops already above 0.5 GeV$^2$, where its incorrect short-distance behavior starts to dominate.

In the lower plot of Fig.~\ref{figKw} we show
the various predictions for the dependence on $x$ of the asymmetric double-virtual TFF $F((1-x)\text{GeV}^2,x\,\text{GeV}^2)/F(0,0)$ at fixed $Q_1^2+Q_2^2=1\,\text{GeV}^2$.
Here the lattice result and the HW1 model result are somewhat above the estimated error band of the dispersive approach of Ref.~\cite{Hoferichter:2018kwz},
whereas the SW model is completely within the latter. The HW2 and SS results are further away but have a shape which is similar to that of the dispersive result.
Note that $x=0$ corresponds to the single-virtual case, where the DRV interpolator is fitting the corresponding experimental data.
There the central value of the dispersive result agrees with the DRV interpolator, while the (central) lattice result appears too high.
Away from $x=0$ the DRV interpolator has a too strong $x$-dependence, which seems to be a reflection of its singular curvature parameter $\hat\beta$
(see Table \ref{IRUVbehavior}).

\section{Hadronic light-by-light contribution to the muon $g-2$}

In order to use the holographic QCD results for the HLBL contribution to $a_\mu$
without having to use the full 3-dimensional integral representation of the relevant two-loop diagram,
Ref.~\cite{Cappiello:2010uy} employed two variants of (extended) DIP interpolators,
\bea
\label{DIP1}
K(Q_1^2,Q_2^2)^{\rm DIP1}&=&1+\lambda\left( \frac{Q_1^2}{Q_1^2+m_1^2}+\frac{Q_2^2}{Q_2^2+m_2^2} \right)
+\sum_{i=1}^2 \eta_i \frac{Q_1^2 Q_2^2}{(Q_1^2+m_i^2)(Q_2^2+m_i^2)}\\
\label{DIP2}
K(Q_1^2,Q_2^2)^{\rm DIP2}&=&1+\sum_{i=1}^2\lambda_i\left( \frac{Q_1^2}{Q_1^2+m_i^2}+\frac{Q_2^2}{Q_2^2+m_i^2} \right)
+ \eta \frac{Q_1^2 Q_2^2}{(Q_1^2+m_1^2)(Q_2^2+m_1^2)}
,
\eea
where $m_2=m_\rho$ and the free parameters were used to include a slope parameter $\hat\alpha=-1.76$ in accordance with the current
world average and also the range obtained by the bottom-up models. Only one of the curvature parameters could be
matched at a time and preference was given to the parametrization obtained by fitting $\hat\beta$ to the average value
of $3.33$ GeV$^{-4}$. Finally, the remaining parameter was fixed by requiring agreement with the LO QCD result $F^\infty$
at $Q_1^2=Q_2^2\to\infty$. The Brodsky-Lepage constraint \cite{Lepage:1979zb,*Lepage:1980fj,*Brodsky:1981rp} could however not be incorporated simultaneously.
The finite limit $K^{\rm DIP1,2}(Q^2\to\infty,0)$ was instead taken as representing a nonzero value of the magnetic susceptibility $\chi_0$
appearing in the short-distance behavior far away from the pion pole \cite{Nyffeler:2009tw}, $F_{\pi^{0*}\gamma^*\gamma^*}(Q^2\to\infty,Q^2\to\infty,0)\to-f_\pi\chi_0/3$.
The latter would be of interest in evaluations beyond the pion-pole approximation  (in particular with the HW1 model, which
has a nonvanishing $\chi_0$ \cite{Gorsky:2009ma}), but here we restrict ourselves to the more well-defined pion-pole contribution.

\begin{table}
\bigskip
\begin{tabular}{l|c|c|c|c|}
\toprule
 Model  & $a_\mu^{\pi^0}$ & $a_\mu^{\eta}$ & $a_\mu^{\eta'}$ & sum \\ 
\colrule
SS  & 4.83 & 1.17 & 0.78$|$0.95 & 6.94 \\
\colrule
HW1 & \hl{6.52} & 1.82 & 1.32$|$1.56 & \hl{9.90} \\
HW2  & 5.66 & 1.48 & 1.03$|$1.24 & 8.37 \\
SW  & 5.92 & 1.59 & 1.12$|$1.34 & 8.85 \\
\colrule
DIP1 & 6.54 & 1.90 & 1.44$|$1.69 & 10.14 \\
DIP2 & 6.58 & 1.92 & 1.46$|$1.71 & 10.22 \\
\colrule
DRV \cite{Danilkin:2019mhd} & 5.6(2) & 1.5(1) & 1.3(1) & 8.4(4) \\
\botrule
\end{tabular}
\caption{Results in multiples of $10^{-10}$ for $f_\pi=92.4 \,\text{MeV}$ for the Sakai-Sugimoto model, three bottom-up models, and the DIP interpolation of
Ref.~\cite{Cappiello:2010uy} based on the curvature $\hat\beta$ averaged over the bottom-up models. The last line refers to the new interpolator (DRV)
of Ref.~\cite{Danilkin:2019mhd} fitted to experimental data including the preliminary ones from BESIII.
Numerical errors in the integrations carried out for the
holographic models are estimated as smaller than $1\cdot 10^{-12}$. For estimating also the contributions $a_\mu^{\eta}$ and $a_\mu^{\eta'}$ we have 
included the experimental masses for $\eta$ and $\eta'$, but kept the results for $K(Q_1^2,Q_2^2)$ obtained 
in the chiral limit and simply rescaled $F(0,0)$ by the central experimental values quoted in \cite{Danilkin:2019mhd}.
The holographic results do not include the experimental errors for the various $F(0,0)$, which in $a_\mu$ amount to
about $\pm 2$\%, 3.5\%, and 4.5\% for $\pi^0$, $\eta$, and $\eta'$, respectively.
However, for $\eta'$ we also give as a second value
the result of the alternative and presumably more realistic extrapolation obtained by additionally upscaling the mass scale within $K(Q_1^2,Q_2^2)$
by +10\% in line with the higher $\Lambda$ parameter in DRV($\eta'$), cf.\ Fig.~\ref{fig18comp}, and use this for the estimated sum total.
}
\label{amuresults}
\end{table}

\begin{figure}
\bigskip
\includegraphics[width=0.65\textwidth]{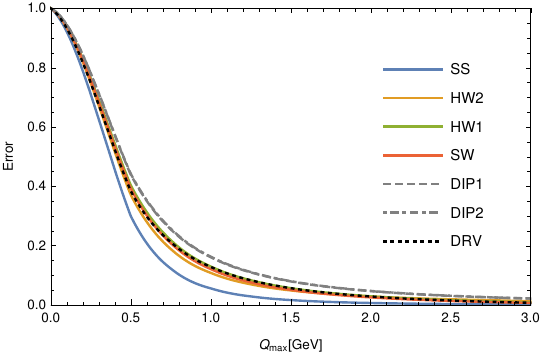}
\caption{Remaining errors in the integral for $a_\mu^{\pi^0}$, when this is cut off at $Q_1^{\rm max}=Q_2^{\rm max}=Q_{\rm max}$ (plotted over $Q_{\rm max}$
instead of $Q_{\rm max}^2$ to spread out the low-$Q$ region).}
\label{figconvergence}
\end{figure}

In Table \ref{amuresults} we show the results of the full 3-dimensional integration that we have carried out
using the formulae of Ref.~\cite{Nyffeler:2016gnb} 
with the holographic results for $K(Q_1^2,Q_2^2)$ (which also require numerical evaluation).
The estimated numerical error of our results is below $1\cdot 10^{-12}$.
In Fig.~\ref{figconvergence} we also display the error of the integrations when these are cut off at some value $Q^{\rm max}$
for both loop momenta, showing the slightly different convergence behavior of the various models.
Evidently, the DIP interpolators do not well represent the results of the bottom-up models but overestimate them
by an amount which is larger than their deviations from each other. This is in line with the discrepancy
displayed in Figs.~\ref{fig3comp} and \ref{fig18comp} between the TFF of the bottom-up models and the DIP interpolators.

Given that the bottom-up holographic models are all in remarkable agreement with the low-energy data for single-virtual pion TFF
as shown in Fig.~\ref{fig3comp}, while also bracketing the lattice results \cite{Gerardin:2019vio} of double-virtual TFF (Fig.~\ref{figKw}),  
they provide, taken together, an arguably strong prediction for $a_\mu$ consisting of
the range \st{$(5.7\ldots 6.1)\cdot 10^{-10}=5.9(2)\cdot 10^{-10}$}\hl{$(5.7\ldots 6.5)\cdot 10^{-10}=6.1(4)\cdot 10^{-10}$}. 
[Notice that this does not include the experimental uncertainty in $F_{\pi^0}(0,0)$, but corresponds to the choice $f_\pi=92.4$ MeV
in (\ref{Fpi00}).]

In Table \ref{amuresults} 
these results are also compared with the result obtained in Ref.~\cite{Danilkin:2019mhd} with a new interpolator (DRV),
fitted to experimental data up to 9 GeV$^2$ for $\pi^0$ including the preliminary ones from BESIII.
This fit for single-virtual pion TTF happens to be indistinguishable from the SW model within
the line thickness in Fig.~\ref{fig3comp}, but for double-virtual TTF there are significant differences as shown in Fig.~\ref{figKw}.
In the latter, the SW results are fully consistent with dispersive results, while this is not the case for the DRV interpolator.
Correspondingly, the SW model yields a somewhat higher value for $a_\mu^{\pi^0}$ than obtained by the DRV interpolator.

Also given are simple extrapolations to $a_\mu^\eta$ and $a_\mu^{\eta'}$ using the same function $K(Q_1^2,Q_2^2)$ 
obtained in all the holographic models in the chiral limit, but rescaling $F(0,0)$ according to
the central experimental values quoted in \cite{Danilkin:2019mhd} ($F_\eta(0,0)/F_{\pi^0}(0,0)=0.2736/0.2725$ and $F_{\eta'}(0,0)/F_{\pi^0}(0,0)=0.3412/0.2725$) and using
the physical masses of $\eta$ and $\eta'$ in the 2-loop integral for $a_\mu^{\eta,\eta'}$.%
\footnote{An approximate evaluation of $a_\mu^{\pi^0,\eta,\eta'}$ has been carried out for the HW1 model generalized to 
finite quark masses in Ref.~\cite{Hong:2009zw},
but using only a finite number of vector meson modes.
The results given in Ref.~\cite{Hong:2009zw} are significantly higher than ours in the chiral limit, even for $a_\mu^{\pi^0}$, although the single-virtual pion TFF for massive pions
obtained there appear to be extremely close to our chiral result.}
(See e.g.\ Ref.~\cite{Masjuan:2017tvw,Guevara:2018rhj} for a discussion of the effects of finite quark masses on pseudoscalar TFFs.)
Judging from the fitting parameter $\Lambda$ obtained in Ref.~\cite{Danilkin:2019mhd}, the bottom-up holographic results for $K(Q_1^2,Q_2^2)$
match also quite well the single-virtual data for $\eta$, but for $\eta'$ a 10\% higher $\Lambda$ was found.
As is shown in Fig.~\ref{fig18comp}, a corresponding rescaling of the mass scales in $K(Q_1^2,Q_2^2)$ produces a very good fit of
double-virtual data for the $\eta'$ TFF in the HW1 and SW cases, which appear to do even better than the DRV interpolator. We have therefore also included the resulting
somewhat higher estimates for $a_\mu^{\eta'}$ in Table \ref{amuresults} (which may give an idea of the uncertainty
in these extrapolations to also cover $\eta'$)
and used them for the estimated sum total.

\section{Conclusion and outlook}

As displayed in Fig.~\ref{fig3comp}, the bottom-up AdS/QCD models considered here and previously in Ref.~\cite{Cappiello:2010uy}
reproduce remarkably well the available experimental data for the single-virtual pion TFF, including the preliminary data from BESIII presented recently in Ref.~\cite{Danilkin:2019mhd}.
The top-down model of Sakai and Sugimoto turns out to have a larger slope parameter but also larger upward curvature compared to the bottom-up models. At the smallest values of $Q^2$ this seems perfectly consistent with the preliminary data from BESIII, but for all $Q^2>0.5 \,\text{GeV}^2$ the experimental data are underestimated, so that the result for $a_\mu^{\pi^0,\rm SS}$ of $4.8\cdot 10^{-10}$ could perhaps be taken as a lower limit to which positive contributions necessary to make contact with the correct short-distance behavior need to be added.
The bottom-up holographic models, on the other hand, reproduce the leading-order asymptotic pQCD behavior either perfectly or to a large percentage, as shown in Table \ref{IRUVbehavior}.
The HW1 model, which asymptotes to 100\% of the asymptotic result, tends to somewhat overestimate the experimental data
as well as the extrapolations to the double virtual case from lattice and dispersive studies. 
The SW model, which saturates at 89\%, is (perhaps fortuitously) in the right ballpark for larger virtualities where
pQCD corrections beyond LO \cite{Melic:2002ij} are still important. It turns out to have a near-perfect agreement with the recent
fits of single-virtual data Ref.~\cite{Danilkin:2019mhd} up to 9 GeV$^2$, and with the recent lattice 
of Ref.~\cite{Gerardin:2019vio} in the symmetric double-virtual case.
The HW2 model appears to underestimate data at large virtualities by a significant amount, but agrees well with
low-energy data (and better than the other models at the lowest $Q^2$ values, see Fig.~\ref{fig3comp});
notice that one
half of the contribution to $a_\mu^{\pi^0}$ comes from the region $Q^2<0.5\,\text{GeV}^2$ (see Fig.~\ref{figconvergence}).

The DIP interpolators used in Ref.~\cite{Cappiello:2010uy} to approximate the bottom-up models appear to severely
overestimate all the results for the pion TFF both with respect to the holographic models as well as the experimental data. Moreover, they fail to include the Brodsky-Lepage constraint \cite{Lepage:1979zb,*Lepage:1980fj,*Brodsky:1981rp}.
From the full evaluation of the pion-pole contribution to $a_\mu$ in the holographic models as carried out here, we arrive at somewhat smaller numbers than presented in Ref.~\cite{Cappiello:2010uy}: instead of
the value $a_\mu^{\pi^0}=6.54(25)\cdot 10^{-10}$ quoted therein, we find that the spread of the bottom-up AdS/QCD results is given by \hl{$6.1(4)\cdot 10^{-10}$}
for fixed $f_\pi=92.4$ MeV. This is in between the central results given by a simple VMD model ($a_\mu^{\rm VMD}=5.7\cdot 10^{-10}$)
and the LMD+V model ($a_\mu^{\rm VMD}=6.3\cdot 10^{-10}$) \cite{Nyffeler:2016gnb}, it is \st{at the lower end of}\hl{slightly smaller but fully consistent with} the result
\cite{Hoferichter:2018kwz} ($a_\mu^{\rm dispersive}=6.3(3)\cdot 10^{-10}$) obtained in the dispersion theory framework of Ref.~\cite{Colangelo:2015ama}. \hl{It also agrees}
\st{but it happens to agree remarkably} well
with the recent lattice result \cite{Gerardin:2019vio} of $5.97(36)\cdot 10^{-10}$ for the pion-pole contribution, \hl{which after data-driven corrections reads}
\st{(prior to data-driven corrections which increase this value to} $6.23(23)\cdot 10^{-10}$.

The bottom-up holographic results are somewhat higher than the result $5.6(2)\cdot 10^{-10}$ obtained with the new (DRV) interpolator introduced in Ref.~\cite{Danilkin:2019mhd},
which at the relevant values of $Q^2$ has a stronger dependence on the asymmetry parameter $w$ (defined in (\ref{fw})) than the holographic result
and also the results from lattice and dispersion relations.
\st{The central value $5.9 \cdot 10^{-10}$ of the holographic results for $a_\mu^{\pi^0}$ is given by the SW model, which for the single-virtual pion TFF}
\hl{The SW model has a single-virtual pion TFF which}
practically coincides with the DRV9 interpolator
used in Ref.~\cite{Danilkin:2019mhd} to account for the new experimental data, while agreeing
 much better with 
dispersive results \cite{Hoferichter:2018kwz} as well as with the currently best lattice results
for the double-virtual pion TFF of Ref.~\cite{Gerardin:2019vio}.
Hence, \st{this} the SW result $5.9 \cdot 10^{-10}$ could already be taken as an improvement of the data-driven estimate of \cite{Danilkin:2019mhd} which included preliminary BESIII data,
increasing the central value by $0.3\cdot 10^{-10}$.

Because the SW and HW1 results bracket experimental data and double virtual extrapolations for the pion TFF
at larger $Q^2$, while at the lowest $Q^2$ the new experimental data summarized by the DRV4 interpolator are closely
bracketed by the SW and HW2 results, the three bottom-up holographic QCD models 
taken together appear to be a very plausible interpolator for estimating
the pion pole contribution, leading to \hl{$a_\mu^{\pi^0}=6.1(4)\cdot 10^{-10}$} (for
fixed $f_\pi=92.4$ MeV, i.e., $F(0,0)=0.274\,\text{GeV}^{-1}$), \st{slightly smaller but overlapping} \hl{in remarkably good agreement}
with the most recent dispersive \cite{Hoferichter:2018kwz} and lattice \cite{Gerardin:2019vio} results.
It will be interesting to see how the various
results for the double-virtual pion TFF compare with future improvements of low-energy experimental or lattice data. 

Given the good agreement of the HW1 results with available data for the pion TFF, it would also be interesting to revisit
the analysis of Ref.~\cite{Hong:2009zw}, where finite quark masses were included in this model, but where $a_\mu^{\pi^0,\eta,\eta'}$ was evaluated
only approximately. This would permit full holographic predictions for the heavier pseudoscalars, for which we have
provided only rough extrapolations in Table \ref{amuresults}.

In future work we plan to consider also glueball-photon couplings as obtained in the SS model and the resulting HLBL contribution to $a_\mu$.
As shown recently in Ref.~\cite{Leutgeb:2019lqu}, in this top-down holographic model the pseudoscalar glueball is predicted to have a large coupling
to vector mesons, as found previously also for scalar and tensor glueballs \cite{Brunner:2015oqa},
and thus also an important coupling to photons \cite{LRtoappear}.
At least for the scalar glueball there is some evidence that the SS model generalized to include finite quark masses
is able to describe the decay pattern of glueballs quantitatively \cite{Brunner:2015yha,Brunner:2015oga}.
The present analysis suggests that the glueball contributions to $a_\mu^\mathrm{HLBL}$ 
would be underestimated by some 20\% due to the incorrect UV behavior of the SS model, but this should be sufficient for a rough estimate.

\begin{acknowledgments}
We thank Massimo Passera and Massimiliano Procura for most useful discussions as well as 
Luigi Cappiello, Oscar Cat\`a, Giancarlo D'Ambrosio, 
Deog Ki Hong, and Doyoun Kim for correspondence.
We are particularly grateful to Igor Danilkin for providing checks of our numerical calculations.
J.~L.\ was supported by the FWF doctoral program
Particles \& Interactions, project no. W1252-N27.
\end{acknowledgments}

\raggedright
\bibliographystyle{JHEP}
\bibliography{hlbl}

\end{document}